# Ferroelectricity in HfO$_2$ from a chemical perspective

Jun-Hui Yuan,[1,2] Ge-Qi Mao,[1,2] Kan-Hao Xue,[1,2*] Na Bai,[1,2] Chengxu Wang,[1,2] Yan Cheng,[3] Hangbing Lyu,[4] Huajun Sun,[1,2] Xingsheng Wang,[1,2] and Xiangshui Miao[1,2]

[1]School of Integrated Circuits, School of Optical and Electronic Information, Huazhong University of Science and Technology, Wuhan 430074, China

[2]Hubei Yangtze Memory Laboratories, Wuhan 430205, China

[3]Key Laboratory of Polar Materials and Devices (MOE), Department of Electronics, East China Normal University, Shanghai 200241, China

[4]Key Laboratory of Microelectronics Devices and Integrated Technology, Institute of Microelectronics, Chinese Academy of Sciences, Beijing 100029, China

*Corresponding Author, e-mail: xkh@hust.edu.cn (K.-H. Xue)

## Abstract

Ferroelectricity observed in thin film HfO$_2$, either doped with Si, Al, etc. or in the Hf$_{0.5}$Zr$_{0.5}$O$_2$ form, has gained great technical significance. However, the soft mode theory faces a difficulty in explaining the origin of such ferroelectricity. In this work, we propose that the 7 cation coordination number of HfO$_2$/ZrO$_2$ lies at the heart of this ferroelectricity, which stems from the proper ionic radii of Hf/Zr compared with O. Among the numerous compounds with non-centrosymmetric nature, e.g., *mm*2 point group, HfO$_2$ and ZrO$_2$ are special in that they are close to the border of 7 and 8 cation coordination, such that the 8-coordination tetragonal intermediate phase could greatly reduce the switching barrier. Other 7-coordination candidates, including SrI$_2$, TaON, YSBr and YOF are also studied in comparison to HfO$_2$/ZrO$_2$, and six switching paths are analyzed in detail for the *Pca*2$_1$ phase. A rule of preferred switching path in terms of ionic radii ratio and coordination number has been established. We also show the possible route from ferroelectric *Pca*2$_1$ phase to monoclinic *P*2$_1$/c phase in HfO$_2$, which is relevant to the fatigue phenomenon.

Ferroelectrics based on hafnia, such as doped $HfO_2$ [1,2] and $Hf_{0.5}Zr_{0.5}O_2$, are quite unusual compared with conventional perovskite ferroelectrics. First of all, though these oxides consist of mixed ionic-covalent bonding, it is unclear whether the covalence should play any role in the ferroelectricity. On the other hand, in perovskite oxide ferroelectrics the existence of orbital hybridization and covalent bonding is important for the ferroelectric distortion [3]. $HfO_2$ even seems to belong to a new ionic ferroelectric family as it is more ionic than covalent. Secondly, a recent study has revealed a flat phonon dispersion in hafnia-ferroelectrics [4], which together with the presence of spacers, implies that collective ionic movement may not be so significant in these ferroelectrics. The ferroelectricity in hafnia-based oxides has been attributed to a polar $Pca2_1$ orthorhombic phase of $HfO_2$ [1,5]. Besides the above special features, another puzzle is that other oxide ferroelectrics in the $Pca2_1$ or similar structures are rarely reported, which renders it difficult to identify the material-specific properties of $Pca2_1$-$HfO_2$ from other members in the ionic ferroelectric family. In principle, however, there are a large number of orthorhombic space groups with mirror symmetry broken in at least one direction (e.g., $mm2$ point group). In order to explore more ionic ferroelectric oxides, it is highly desirable to understand the origin of such $Pca2_1$ phase in $HfO_2$, and to propose some guide for materials design in the ionic ferroelectric oxide family. In addition, this also provokes a serious concern, for what is the exact material gene of $HfO_2$/$ZrO_2$ for being the special ferroelectric material at the nanoscale.

$HfO_2$ adopts a cubic fluorite structure at temperatures above 2773 K [6] and below its melting point. Nevertheless, such structure is dynamically unstable at low temperatures, and in **Figure 1** we demonstrate the phonon spectra of cubic $HfO_2$ calculated at zero temperature using density functional theory. The Vienna *Ab initio* Simulation Package [7,8] was implemented in our calculations, with the projector augmented wave method and the plane wave kinetic energy cutoff was 600 eV. Detailed parameters for calculation are listed in **Supplementary Note 1**. The strong imaginary mode near the X-point indicates the dynamical instability. This imaginary mode would yield a cubic-to-tetragonal phase transition at low temperature, and our calculation shows that tetragonal $HfO_2$, as well as its derivatives including $P2_1/c$ and $Pca2_1$ phases, are all free of imaginary phonon modes (see **Figure S1**). This is consistent with numerous previous results[4,9] and in a bit contradiction to the work of Huan *et al.* [10] who showed some imaginary phonon mode for the tetragonal phase. While tetragonal $HfO_2$ is dynamically stable, it is well-known that the ground state phase of $HfO_2$ is monoclinic $P2_1/c$ [11,12]. By transforming the Hf cations from VIII-coordination (8C for short, and the same notation will be used hereafter) as in tetragonal $HfO_2$ to 7C as in monoclinic $HfO_2$, the overall energy is lowered. Indeed, the ionic radius of Hf is not as large as Ce, which explains why in the low temperature phase $HfO_2$ each Hf cation is only surrounded by 7 O anions (see the circled Hf cations in **Figure 1**), rather than 8 as in the case of fluorite $CeO_2$ [13]. With the same argument, we find that instead of a soft phonon mode at the zone center, the emergence of $Pca2_1$ $HfO_2$ is driven by the need of 8C to 7C chemical configuration for Hf as shown in **Figure 1**. The discrepancy lies in that, to obtain the $P2_1/c$ phase, the two oxygen anions within the ellipse move in opposite directions to yield a symmetric structure; but in deriving the $Pca2_1$ phase, the two oxygen anions within the ellipse move in the same direction and therefore mirror symmetry is broken along its *c*-axis. These two O anions become 3C after distortion (here named $O_{III}$), while the 4C O anions are named $O_{IV}$ in the $Pca2_1$ phase. From $P4_2/nmc$ to $Pca2_1$, a half of the $O_{IV}$ anions suffer from a slight movement along the *b*-axis, which is however similar to the $P2_1/c$-distortion. In $Pca2_1$ $HfO_2$, the ferroelectric switching is mainly caused by the movement of $O_{III}$ anions, while the $O_{IV}$ anions can be regarded as fixed. A potential benefit of the $Pca2_1$ distortion lies in that the unit cell does not have to expand greatly like in the $P2_1/c$ case [14].

Traditional soft-mode theory [15] does not explain the origin of ferroelectricity in $HfO_2$, since there is no paraelectric mother phase of $HfO_2$ that has been reported to possess a soft mode at the Brillouin zone center. The soft mode found in the phonon spectra of cubic $HfO_2$ (**Figure 1**) lies on the zone border and yields the cubic-to-tetragonal phase transition, which may only yield antiferroelectric phases. On account of the fact that $Pca2_1$-$HfO_2$ is not thermodynamically stable, the ferroelectricity in hafnia is more related to the energy minimization at the nanoscale rather than special phonon properties. A chemical perspective regarding the preference of 7C for Hf in $HfO_2$ may provide a more useful theoretical explanation than the soft-mode theory.

Ionic compounds are signified by relatively large coordination numbers. A qualitative argument for the structure prediction in cubic AB ionic compounds exists [16], by defining the radius ratio

$$\rho = \frac{r_{small}}{r_{big}}, \ 0 < \rho \leq 1$$

As $\rho$ increases, the cubic structure evolves from zinc blende (4C) to rock salt (6C), and finally to the CsCl structure (8C). The two critical $\rho$ values are estimated to be 0.326 and 0.717 by Michmerhuizen *et al.* [17]. However, 5C and 7C are very rare and we are not aware of any AB ionic compound of these categories. For $AB_2$-type ionic compounds, the focus should be put on the cation coordination number, and

$$\rho = \frac{r_{cation}}{r_{anion}}$$

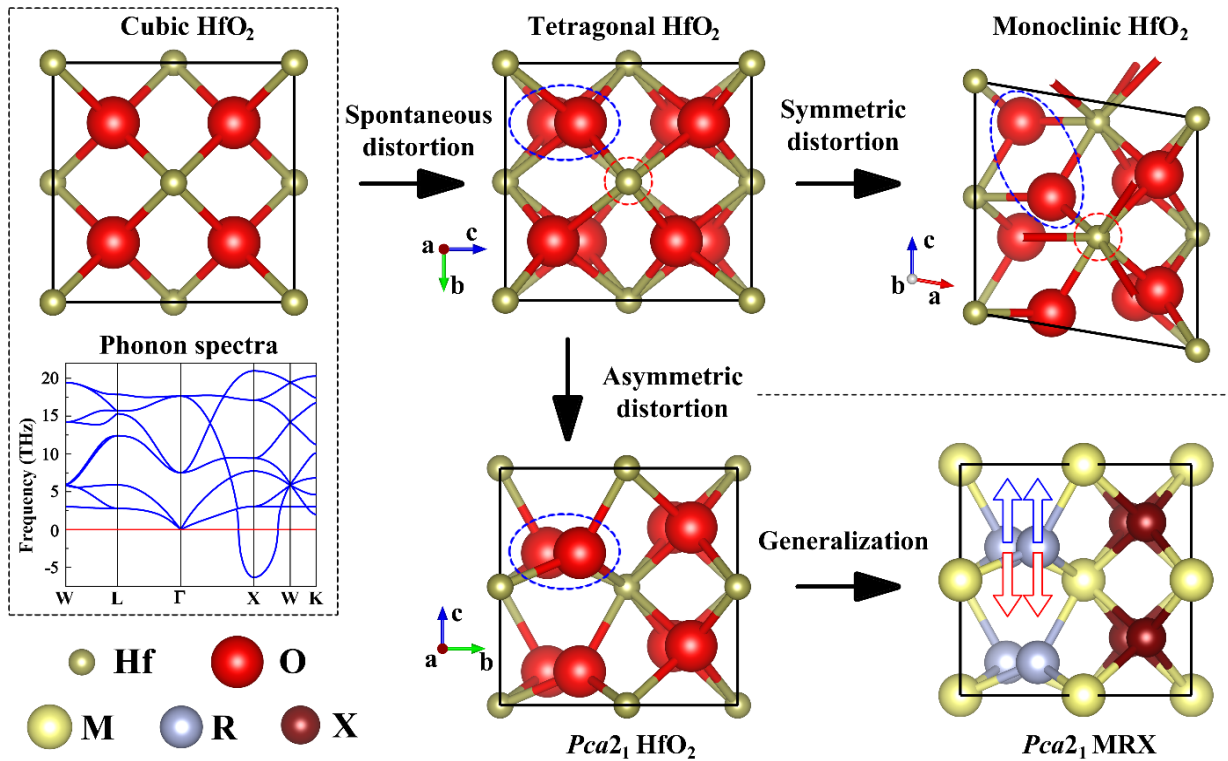

**Figure 1.** Crystal structures of cubic fluorite $HfO_2$, tetragonal $P4_2/nmc$ $HfO_2$, monoclinic $P2_1/c$ $HfO_2$ and orthorhombic $Pca2_1$ $HfO_2$, with their relations. The predicted $Pca2_1$ MRX is further derived from $Pca2_1$ $HfO_2$, where internal and external switching paths are marked by blue and red arrows, respectively.

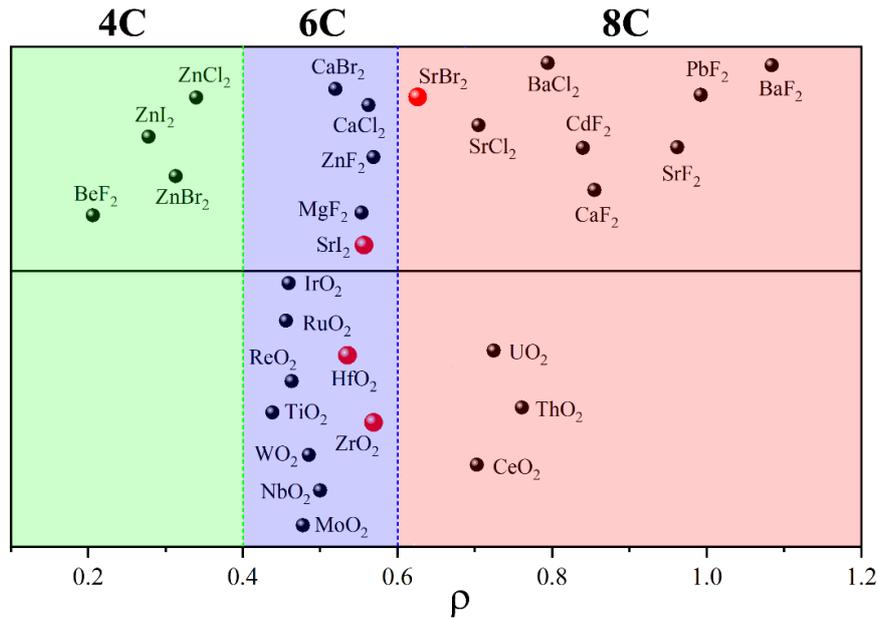

**Figure 2.** Relationship between cation coordination number and cation/anion radius ratio for AB$_2$ compounds. Cations with +2 and +4 valences are separated vertically. The 7C compounds are specially marked by red balls.

For the same ρ value, cation coordination for AB$_2$ is generally larger than that of AB. For example, the 8-coordination CeO$_2$ (fluorite structure) only possesses a ρ value of 0.703. Indeed, the anion coordination number is around half that of cation's in AB$_2$. As ρ increases, AB$_2$ ionic compounds typically evolves from 6C (like rutile and anatase), 7C (typically baddeleyite), 8C (typically fluorite) to 9C (like cotunnite, but this structure may also be regarded as 7C if neglecting the two longer bonds). In **Figure 2** we list some typical examples to show the influence of ρ on the AB$_2$ cation coordination, by simply setting 0.4 and 0.6 as the threshold values between 4C, 6C and 8C. Odd-coordination number compounds are specially marked in red color. Although this is a very rough prediction method, it makes sense in terms of the general trend. The 4C compounds mostly involve the small cation Be$^{2+}$ or Zn$^{2+}$, while 5C is extremely rare in ionic compounds. And HfO$_2$ and ZrO$_2$ also belong to a rather small group of 7C compounds. The odd-number of anions surrounding the cation allows for the room of structural transition from one stable/meta-stable state to the other stable/meta-stable state. Therefore, 7C can be regarded as a characteristic of ionic ferroelectrics.

As a first application, the 7C theory can help predict possible $Pca2_1$ phases in ionic compounds (see **Supplementary Note 2**). For instance, the ionic radius ratio of $Sr^{2+}$ and $I^-$ also lies close to the boundary between 6C and 8C, possibly yielding a 7C configuration. Its experimental single crystal form is in a $Pbca$ phase (here named $Pbca$-I, as illustrated in **Figure S2(b)**), while the polarized $Pca2_1$ phase of $SrI_2$ is predicted to be isomorphic to $Pca2_1$-$HfO_2$, whose dynamic stability has been confirmed according to phonon spectra calculation (*cf*. **Supplementary Note 2**). In addition, by assigning the $O_{III}$ and $O_{IV}$ sites to different elements, much more $Pca2_1$-phase materials can be predicted. Simply replacing the $O_{IV}$ atoms by N, while substituting +5 valency metals for Hf yields TaON in the $Pca2_1$-structure. YSBr and YOF are examples with Y cations replacing Hf cations. In the $Pca2_1$ phase, $SrI_2$, TaON, YSBr and YOF are all confirmed to be dynamically stable, and their thermodynamic stabilities with respect to $P2_1/c$ and $Pbca$-I) phases are given in **Supplementary Note 2**. Among the phases under investigation, the lowest energy one is $Pnma$ for $SrI_2$, but experimentally obtained single crystal $SrI_2$ or Eu-doped $SrI_2$ single crystal with 7C $Pbca$-I phase has also been synthesized [18]. YSBr also favors a 7C $Pbca$-I structure, which is more stable than $P2_1/c$. For the other compounds, the lowest energy phase is $P2_1/c$ with 7C. It is discovered that all their $Pbca$ or $Pca2_1$ phases are noticeably smaller (~2% to 7%) than the corresponding $P2_1/c$ phase (see **Table S1**), thus mechanical confinement may be used to yield the desired orthorhombic phases, while the $Pca2_1$ phase may further be derived through electrical polarization from the $Pbca$-I phase. Volume confinement is thus a common feature for obtaining the $Pca2_1$-phase against the $P2_1/c$ phase, not limited to $HfO_2$. These materials can theoretically enlarge the research scope of $Pca2_1$-ferroelectrics, but they may not possess ferroelectricity unless the coercive field is less than the dielectric breakdown field. Therefore, it is of great importance to study the migration barriers during polarization switching.

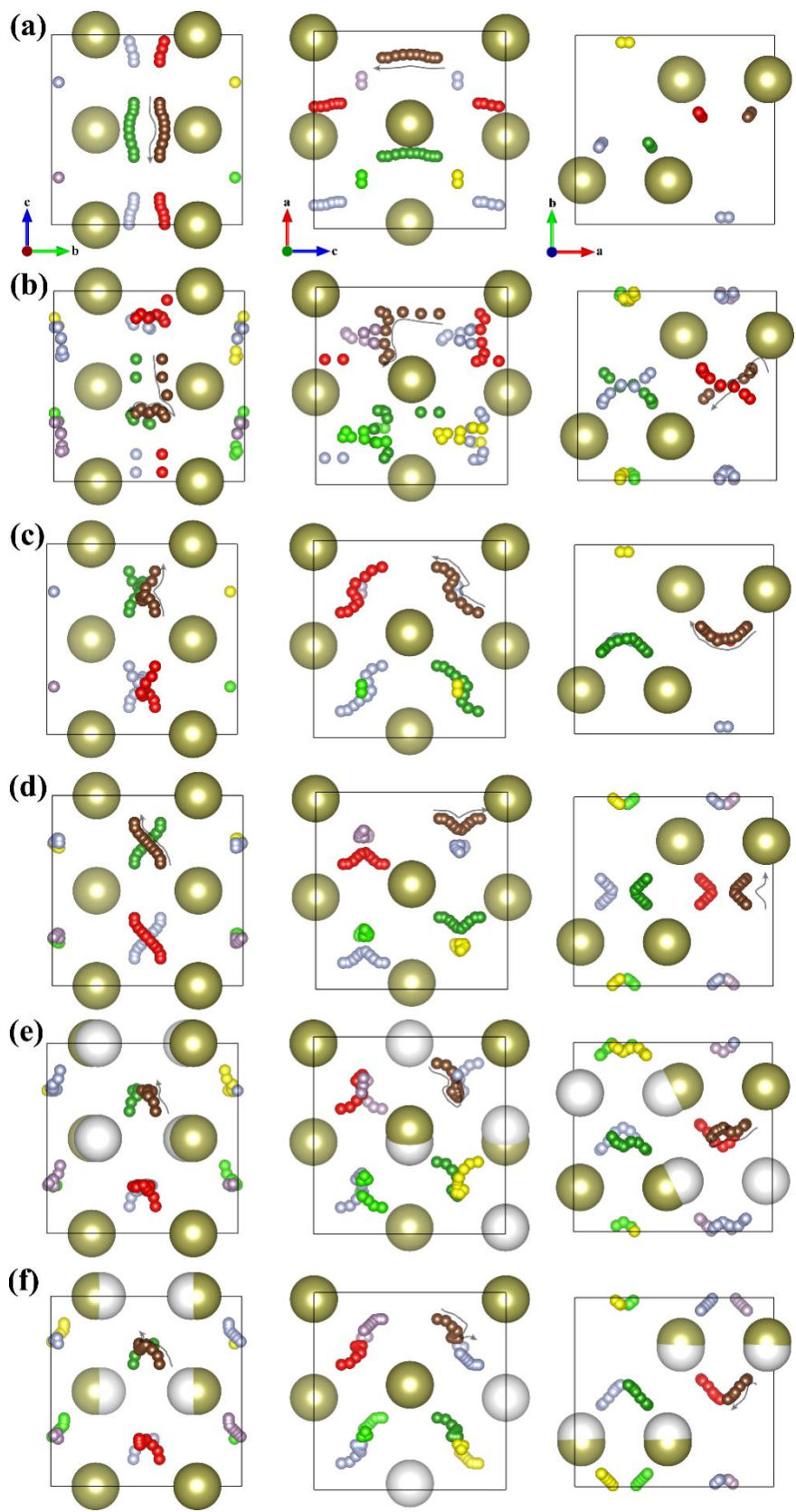

**Figure 3.** Illustration of all six switching paths in $Pca2_1$-HfO$_2$, viewed from three perspectives. The arrows indicate the direction of motion for a selected O atom.

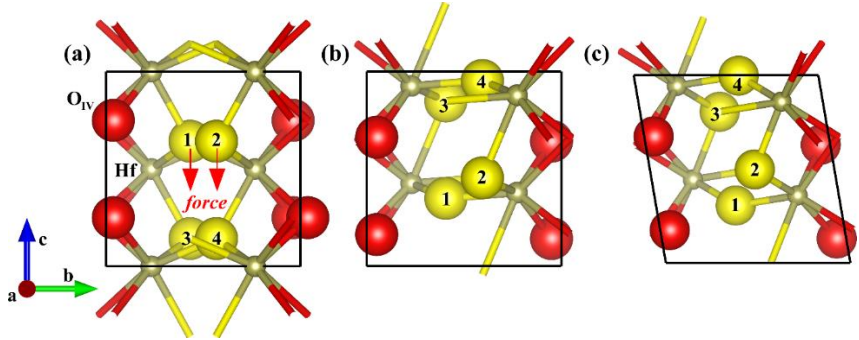

**Figure 4.** Steps from *Pca*2$_1$-phase to *P*2$_1$/*c* phase. (a) Continued high applied electric field on a switched unit cell; (b) switching through the external path with one O$_{III}$ anion (#1) moving ahead of the other (#2); (c) unit cell distortion into the *P*2$_1$/*c* phase. The O$_{III}$ and O$_{IV}$ anions are denoted by yellow and red balls, respectively.

There is rarely ambiguity for the migration path in traditional ferroelectrics, e. g., the fundamental switching modes of PbTiO$_3$ and SrBi$_2$Ta$_2$O$_9$ are illustrated in **Figure S3**. The ferroelectric distortion is caused by the need of covalent Ti-O bonding in PbTiO$_3$ and the lattice mismatch between [Bi$_2$O$_2$]$^{2+}$ and the pseudo-perovskite layers in SrBi$_2$Ta$_2$O$_9$. In the prototype *Pca*2$_1$-phase, however, possible switching paths can be divided into internal paths (left of **Figure S3(c)**) and external paths (right of **Figure S3(c)**), with different spontaneous polarization values. Detailed analysis reveals that the switching may or may not accompany of an exchange of 3C and 4C anions. Besides, the switching may yield the interchange of 3C anions along *b*-axis (crossing), or no crossing along *b*-axis. However, the III-IV exchange is impossible for external paths, therefore the remaining six possible paths (see **Figure S4**) are listed in **Table I**. Take *Pca*2$_1$-HfO$_2$ as an example. **Figure 3** illustrates the detailed atomic movements for all six paths, viewed along the three axes respectively (for others see **Figure S5** to **S9**). The climbing image nudged elastic band (CI-NEB) calculation results for the switching in HfO$_2$ is demonstrated in **Figure S10 (a)**, where the energy barriers for Path-4 and Path-6 are almost identical, both superior to other paths. For other compounds except for ZrO$_2$, the favorite paths are however different. **Table II** specifies the calculated migration barriers during polarization switching in these six compounds for all six paths.

**Table I.** All possible switching paths for HfO$_2$-like $Pca2_1$ ferroelectrics.

| Type | No III-IV Exchange | | III-IV Exchange | |
| --- | --- | --- | --- | --- |
| | No crossing | Crossing | No crossing | Crossing |
| External Path | Path-1: $Pbcm$ | Path-2: $P2_1/c$ | × | × |
| Internal Path | Path-3: $P4_2/nmc$-1 | Path-4: $P4_2/nmc$-2 | Path-5: Exch-1 | Path-6: Exch-2 |

**Table II.** Calculated switching barriers and spontaneous polarizations for all paths in the six compounds under investigation. The underlined values are preferred ones.

| Material | Switching barrier (eV) | | | | | | $P_s$ (μC/cm$^2$) | |
| --- | --- | --- | --- | --- | --- | --- | --- | --- |
| | External | | Internal | | | | External | Internal |
| | Path-1 | Path-2 | Path-3 | Path-4 | Path-5 | Path-6 | | |
| HfO$_2$ | 0.788 | 0.705 | 0.660 | <u>0.3375</u> | 0.700 | <u>0.3383</u> | 69 | <u>52</u> |
| ZrO$_2$ | 0.630 | 0.593 | 0.472 | <u>0.1750</u> | 0.468 | <u>0.1754</u> | 67 | <u>50</u> |
| SrI$_2$ | 0.299 | 0.395 | 0.395 | 0.372 | <u>0.224</u> | 0.302 | 13 | <u>10</u> |
| TaON | <u>0.421</u> | 0.876 | 1.545 | 0.838 | × | × | <u>52</u> | 75 |
| YSBr | <u>0.125</u> | 1.415 | 1.304 | 0.811 | × | × | <u>17</u> | 19 |
| YOF | 0.262 | 0.262 | 0.211 | <u>0.153</u> | × | × | 29 | <u>19</u> |

In HfO$_2$ and ZrO$_2$, the switching is achieved through one or few intermediate phases. For example, Path-1 involves a symmetric $Pbcm$ phase as illustrated in **Figure S2(a)**, while Path-2 to Path-6 all transit via an intermediate structure similar to the well-known tetragonal $P4_2/nmc$ phase (Path-2 is more complicated as discussed later). This is the most significant feature of the polarization switching in $Pca2_1$-HfO$_2$, as revealed by many previous works [10,19–21]. The switching must accompany the breaking of at least one Hf-O bond, therefore the coordination number of Hf either changes as 7-8-7 (forming a new bond first) or 7-6-7 (breaking an old bond first). Obviously, all energetically favorable paths for HfO$_2$ prefers forming the 8C intermediate phase (i.e., Hf is in "7.5C"). This is because tetragonal HfO$_2$ (8C) is energetically much favorable than 6C phases (rutile, anatase, etc.). The same barrier heights for Path-4 and Path-6 reflects little synergic effect between $O_{III}$ and $O_{IV}$ movements in $HfO_2$. Indeed, after the formation of intermediate $P4_2/nmc$ phase by $O_{III}$-to-$O_{IV}$ movement, the subsequent movement may be the original $O_{III}$ (Path-4) or the original $O_{IV}$ (Path-6).

Of the six paths, Path-2 has not been discussed in the literature. Actually, Path-2 is even more complicated as to involve two intermediate phases similar to $P2_1/c$ and $P4_2/nmc$, respectively. For $HfO_2$ or $ZrO_2$, Path-2 possesses a lower barrier value than Path-1. It is noteworthy that a high applied electric field may trigger Path-2 aside from Path-4/Path-6 in $HfO_2$. This not only leads to an unexpected larger polarization and two-step polarization switching, but highly risks transforming the local grain into the monoclinic $P2_1/c$ phase, leading to ferroelectric fatigue. We took the first intermediate phase that was confined to be orthorhombic by the CI-NEB setting, and relaxed it without constraints, which directly yielded the monoclinic phase. The degradation mechanism under high applied electric field is illustrated in **Figure 4**. The driving force for the movement of one $O_{III}$ ion, instead of two simultaneously, lies in maintaining the 7C configuration instead of becoming 6C.

$SrI_2$ prefers the III-VI exchange path and the favorite Path-5 comes through a *Pbca* intermediate phase (here named *Pbca*-III, see **Figure S2(d)**), which is however different from the standard *Pbca*–I phase for $HfO_2$ and the phase discovered by Du *et al.* (*Pbca*–II) [22]. Rather, here the Sr cation is 6C, and it is because bonding breaking occurs ahead of new bond formation that such *Pbca*–III phase emerges. Therefore, $Sr^{2+}$ seems to be "6.5C" in $SrI_2$, distinct from the $HfO_2$ case. The two representative ternary compounds, TaON and YSBr, both prefer Path-1 through a *Pbcm* intermediate phase because any III-VI exchange path is impossible as it leads to a final state with too high energy. As Ta cation is smaller than Hf cation, and S/Br anions are larger than O anions, these render the 6C *Pbcm* intermediate phase favorable. However, ternary $Pca2_1$ compounds not always prefer the external path, since the favorable path for YOF is Path-4.

After setting up the most favorable paths for the six compounds, we link them to the corresponding spontaneous polarization $P_s$ values. Since internal and external paths should in general lead to different spontaneous polarizations, it is only the identification of the actual path that could fulfill this goal unambiguously. The most striking feature is that all compounds only take a favorable switching path with lower $P_s$ value. A comparison between YSBr and YOF is particularly helpful. On the one hand, the smaller anion radii in YOF lead to the internal path with 8C intermediate phase, which further verifies the theory of ionic radius ratio and its impact upon the preferred switching path. On the other hand, though the two materials adopt distinct paths, both of them select the path with the smaller $P_s$ value, in consistent with the discovery that the lower $P_s$ paths are energetically more favorable than the higher $P_s$ paths. Due to the preference of smaller polarization, it is desirable to have more balanced external and internal paths in order to obtain a larger $P_s$, such as in the case of $HfO_2$ (69 μC/cm$^2$ vs. 52 μC/cm$^2$) and $ZrO_2$ (67 μC/cm$^2$ vs. 50 μC/cm$^2$).

We now digress into why $HfO_2$ and $ZrO_2$ fall into the 7C family. Their cation ionic radii ($Hf^{4+}$: 76 pm; $Zr^{4+}$: 78 pm) are apparently too small as to fit the 7C category, if counted from $Hf^{4+}/Zr^{4+}$ and $O^{2-}$ configurations. Moreover, the switching barriers in $Pca2_1$-$ZrO_2$ are substantially lower than that of $Pca2_1$-$HfO_2$. This is at first surprising since the ionic radii of $Zr^{4+}$ is only slightly larger than $Hf^{4+}$ under 7C, by ~2.6%. Although Hf-O bond is stronger than Zr-O bond, this puzzle can be resolved from the radius consideration. The Zr-O and Hf-O bonding are not totally ionic, but still containing a substantial covalent portion. Hence, the amounts of charges carried by Zr and O in $ZrO_2$ are supposed to be lower than that of Hf and O in $HfO_2$, since $ZrO_2$ is more covalent than $HfO_2$. Our Bader charge analysis shows that the average charges on Zr/O and Hf/O are 2.72 e/-1.36 e and 2.86 e/-1.43 e in $ZrO_2$ and $HfO_2$, respectively. Consequently, $Zr^{4+}$ is larger than expected while $O^{2-}$ is smaller than expected in $ZrO_2$, thus the movement of smaller anions within the larger cation lattice is of course much easier. In other words, the substation of Zr for Hf could not only suppress the formation of monoclinic phase, but the coercive fields get substantially reduced to help obtain the ferroelectric properties.

In summary, to explore the ferroelectric material gene of $HfO_2$, we have enlarged the sample set by predicting more possible candidates in the $Pca2_1$-phase. The following conclusions are drawn.

1. Accepting the tetragonal $P4_2/nmc$ phase is a proper mother phase for hafnia-ferroelectrics, the $Pca2_1$ ferroelectric distortion does not stem from a softing of phonon modes. Rather, it is due to the reduction of cation coordination number from 8 (in $P4_2/nmc$) to 7, since the radius of Hf (or Zr) cation is not sufficiently large.

2. The emergence of such ferroelectricity in ionic compounds depends upon a low switching barrier. If the cation/anion radius ratio exactly favors 7C, the switching can be difficult. However, as the ratio is close to 8 or 6, though still falling into the 7C family, the switching can be achieved through an 8C or 6C intermediate phase to reduce the barrier height. Maintaining the ferroelectricity relies on the competition between effective coercive field and dielectric breakdown field.

3. While the partial covalent character in Hf-O and Zr-O bonds could enhance the cation/anion radius ratio to enable 7C, covalent bonding is not required for the $Pca2_1$-ferroelectricity as discussed in this work. Rather, a purely ionic 7C compound may still be potentially ferroelectric as long as (i) its ionic radius ratio favors 7C; (ii) its switching barrier is sufficiently small. This is different from traditional perovskite ferroelectrics where the covalence is indispensable.

4. As Zr is less ionic than Hf, the radius ratio Zr/O in $ZrO_2$ is substantially larger than that of Hf/O in $HfO_2$. Hence, substitution of Zr for Hf in $Pca2_1$-$HfO_2$ may not only suppress the formation of monoclinic phase, but it effectively reduces the ferroelectric switching barrier.

5. The favorite switching path in $Pca2_1$-$HfO_2$ is through an internal path via a $P4_2/nmc$–like intermediate phase as already pointed out in many previous works. However, large applied voltage may further trigger a secondary external switching, leading to enhanced polarization and possible phase transition to the monoclinic $P2_1/c$ phase. This may explain the emergence of irreducible ferroelectric fatigue in hafnia ferroelectrics.

**Acknowledgement**

This work was supported by the National Natural Science Foundation of China under Grant No. 61974049.

# Supplementary Information for

# Ferroelectricity in HfO$_2$ from a chemical perspective

Jun-Hui Yuan,[1,2] Ge-Qi Mao,[1,2] Kan-Hao Xue,[1,2*] Na Bai,[1,2] Chengxu Wang,[1,2] Yan Cheng,[3] Hangbing Lyu,[4] Huajun Sun,[1,2] Xingsheng Wang,[1,2] and Xiangshui Miao[1,2]

[1]School of Integrated Circuits, School of Optical and Electronic Information, Huazhong University of Science and Technology, Wuhan 430074, China

[2]Hubei Yangtze Memory Laboratories, Wuhan 430205, China

[3]Key Laboratory of Polar Materials and Devices (MOE), Department of Electronics, East China Normal University, Shanghai 200241, China

[4]Key Laboratory of Microelectronics Devices and Integrated Technology, Institute of Microelectronics, Chinese Academy of Sciences, Beijing 100029, China

**Corresponding Author**
*E-mail: xkh@hust.edu.cn (K.-H. Xue)

**Note 1. Computational method and parameter settings**

In our calculations, the projector augmented-wave (PAW) method [1,2] and pseudopotentials were adopted, using the Vienna *Ab initio* Simulation Package (VASP 5.4.4). [3,4] The valence electron configurations were: 2s and 2p for O/N/F; 3s and 3p for S; 4s and 4p for Br; 5s and 5p for I; 4s, 4p and 5s for Sr; 4s, 4p, 4d and 5s for Y/Zr; 5p, 5d and 6s for Hf; 5d and 6s for Ta. The plane wave basis was truncated at a 600 eV kinetic energy cutoff. The Perdew-Burke-Ernzerhof (PBE) functional [5] was implemented for the exchange-correlation energy. In structural optimization, a strict 0.001 eV/Å force criterion was used. We employed the following Monkhorst-Pack *k*-point grids for Brillouin zone sampling during geometry optimization: 8×8×8 for *P*2$_1$/*c* and *Pca*2$_1$ phases, 4×8×8 for *Pbca*-I phases. The phonon dispersion was calculated jointly using the VASP code and the PHONOPY code based on the density functional perturbation theory. [6] The ion diffusion barriers were calculated using the climbing image nudged elastic band (CI-NEB) method [7]. In the CI-NEB calculation, the Hellman-Feynman forces on each atom were set to less than 0.03 eV Å$^{-1}$ in all directions, and the energy convergence was below 10$^{-6}$ eV energy difference between two consecutive self-consistent steps. The Berry phase method [8] was used to calculate the spontaneous polarization in *Pca*2$_1$-ferroelectrics (HfO$_2$, SrI$_2$, TaON, YSBr and YOF) along the *c*-axis.

**Note 2. Crystal structures, lattice constants and stabilities of SrI$_2$, TaON, YSBr and YOF**

In **Figure N1**, we have plotted the optimized crystal structures of SrI$_2$, TaON, YSBr and YOF in the *Pca*2$_1$ phase. The corresponding lattice constants are summarized in **Table N1**. For TaON, the oxygen atoms are in 3C, and nitrogen atoms are in 4C. In YSBr and YOF, the halogen elements Br and F atoms are in 3C and the chalcogen elements S and O are in 4C. Furthermore, the dynamic stability of these four *Pca*2$_1$ phases have been confirmed by the phonon dispersions, as shown in **Figure N2**. **Table N2** summarizes the cohesive energy (eV/atom) of six compounds in *P*2$_1$/*c*, *Pbca*-I and *Pca*2$_1$ phases. For HfO$_2$, ZrO$_2$, SrI$_2$, TaON and YOF, the *P*2$_1$/*c* phase has the highest cohesive energy among the three. For YSBr, compared with the other two phases, the *Pbca*-I phase has a higher cohesive energy. In addition, the cohesive energy of *Pbca*-I phase is slightly higher than that of *Pca*2$_1$ phase for all six compounds, but the energy difference is very small (~1 meV/atom to 4 meV/atom).

**Table N1.** The lattice constants of SrI$_2$, TaON, YSBr and YOF in the *Pca*2$_1$ phase

| Materials | *a* (Å) | *b* (Å) | *c* (Å) |
|---|---|---|---|
| SrI$_2$ | 8.2885 | 7.9386 | 8.0524 |
| TaON | 5.1556 | 4.9072 | 5.0336 |
| YSBr | 5.5609 | 5.5076 | 5.5005 |
| YOF | 6.8964 | 6.7061 | 6.7684 |

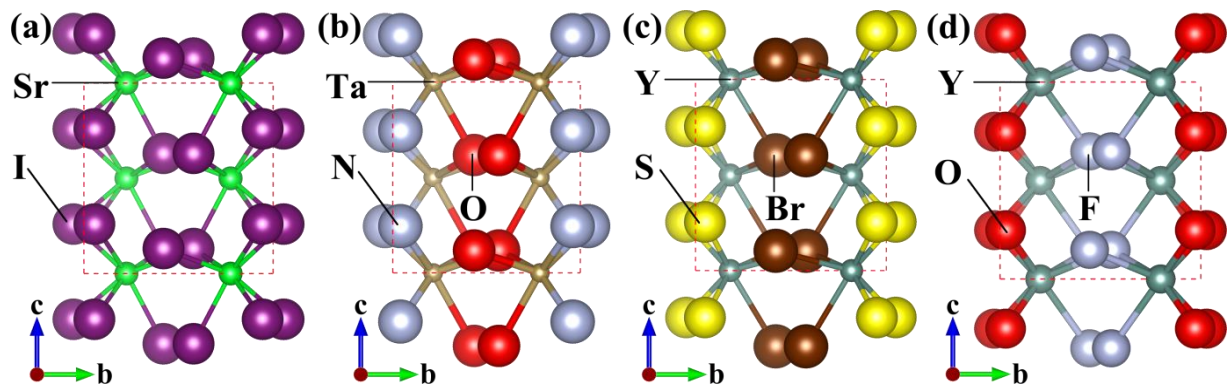

**Figure N1.** The optimized crystal structures of (a) $SrI_2$, (b) TaON, (c) YSBr and (d) YOF in the polar $Pca2_1$ phase.

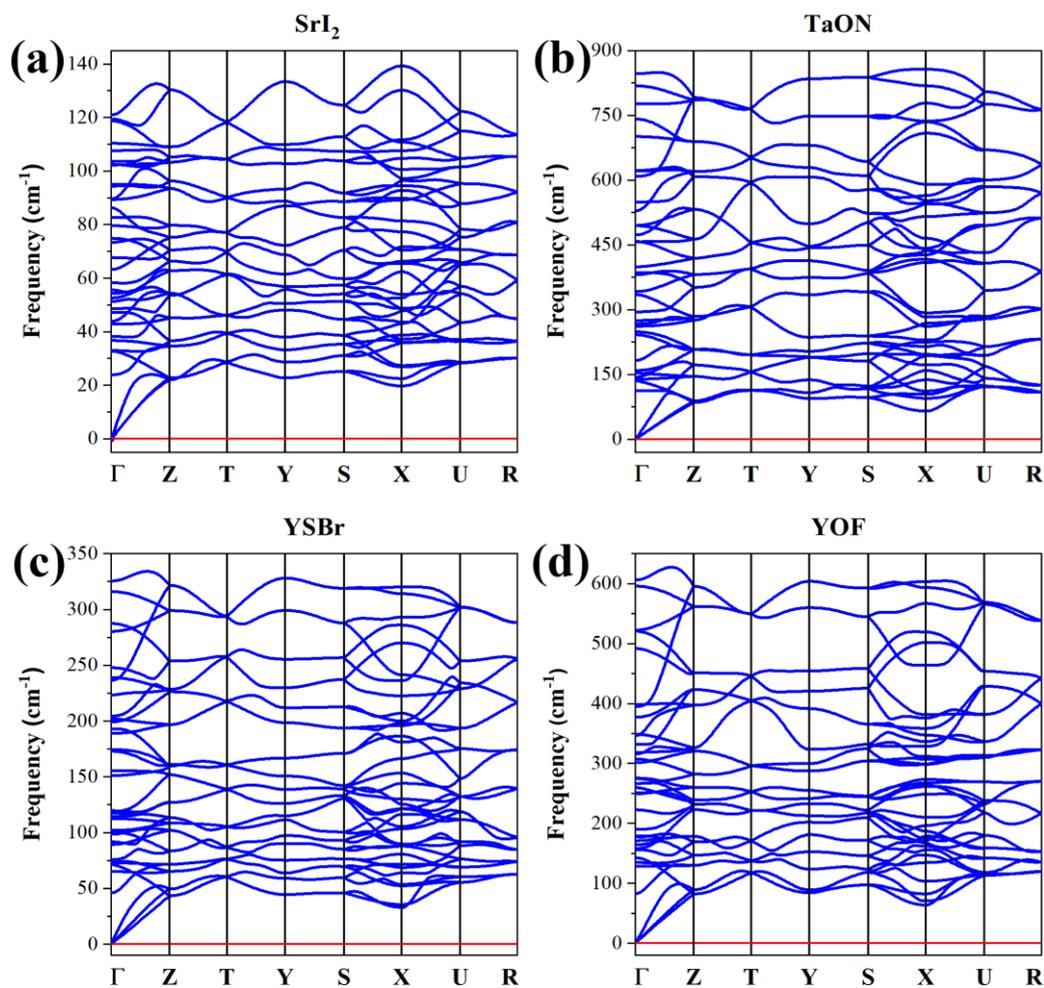

**Figure N2.** Calculated phonon dispersions of (a) $SrI_2$, (b) TaON, (c) YSBr and (d) YOF in the $Pca2_1$ phase.

**Table N2.** The cohesive energies of six compounds in $P2_1/c$, $Pbca$-I and $Pca2_1$ phases, with respect to isolated atoms. The relatively stable phases are underlined.

| Material | Cohesive energy (eV/atom) | | |
| --- | --- | --- | --- |
| | $P2_1/c$ | $Pbca$-I | $Pca2_1$ |
| HfO$_2$ | <u>7.985</u> | 7.960 | 7.957 |
| ZrO$_2$ | <u>7.822</u> | 7.799 | 7.798 |
| SrI$_2$ | <u>3.132</u> | 3.129 | 3.125 |
| TaON | <u>7.505</u> | 7.444 | 7.443 |
| YSBr | 4.872 | <u>4.985</u> | 4.981 |
| YOF | <u>6.943</u> | 6.933 | 6.933 |

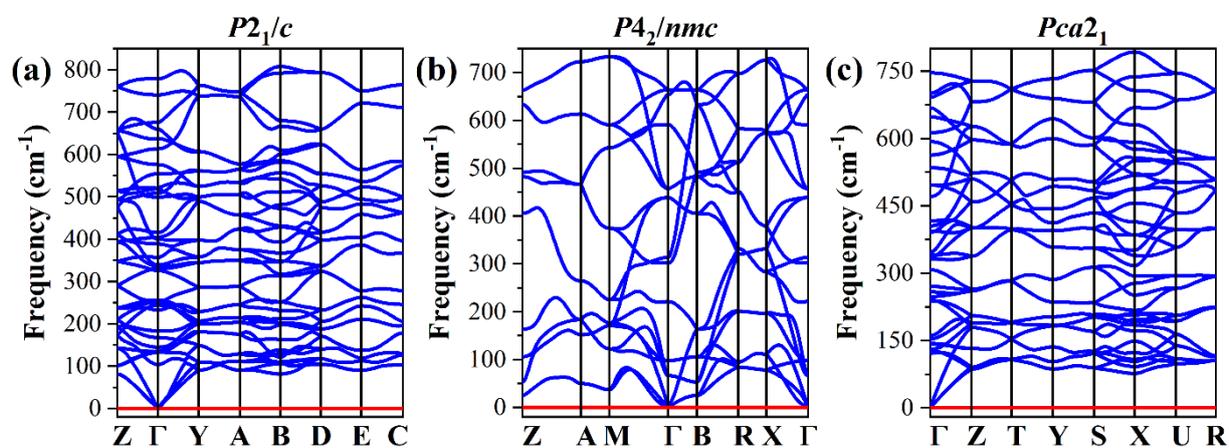

**Figure S1.** Calculated phonon dispersions of HfO$_2$ in (a) $P2_1/c$, (b) $P4_2/nmc$ and (c) $Pca2_1$ phases.

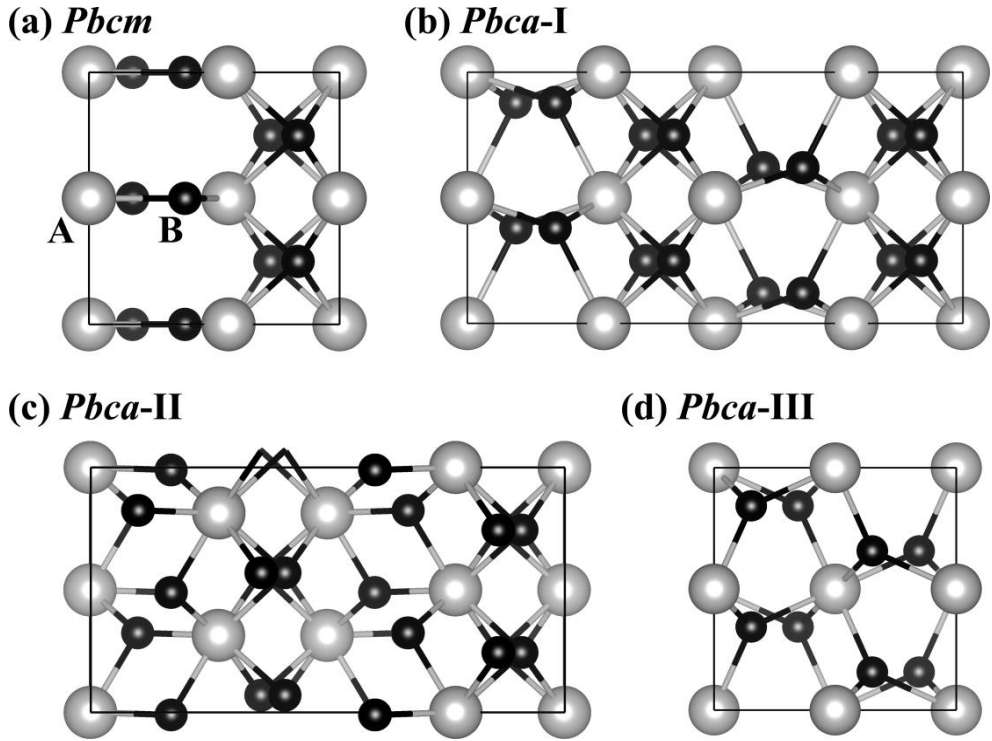

**Figure S2.** The corresponding AB$_2$ crystal structures of (a) *Pbcm* phase, (b) *Pbca*-I phase, (c) *Pbca*-II phase and (d) *Pbca*-III phase.

**Table S1.** The volume expansion ratio (%) of *P2$_1$/c*, *Pbca*-I and *Pca2$_1$* phases in six compounds. Here we take the volume of *Pbca*-I phase as the reference (0% expansion).

| Materials | phases | ratio (%) |
|---|---|---|
| HfO$_2$ | *P2$_1$/c* | 4.508 |
|  | *Pbca*-I | 0 |
|  | *Pca2$_1$* | 0.784 |
| ZrO$_2$ | *P2$_1$/c* | 4.582 |
|  | *Pbca*-I | 0 |
|  | *Pca2$_1$* | 0.884 |
| SrI$_2$ | *P2$_1$/c* | 2.394 |
|  | *Pbca*-I | 0 |
|  | *Pca2$_1$* | 1.089 |
| TaON | *P2$_1$/c* | 2.934 |
|  | *Pbca*-I | 0 |

| | $Pca2_1$ | 0.370 |
|---|---|---|
| YSBr | $P2_1/c$ | 7.260 |
| | $Pbca$-I | 0 |
| | $Pca2_1$ | 0.330 |
| YOF | $P2_1/c$ | 2.836 |
| | $Pbca$-I | 0 |
| | $Pca2_1$ | 0.068 |

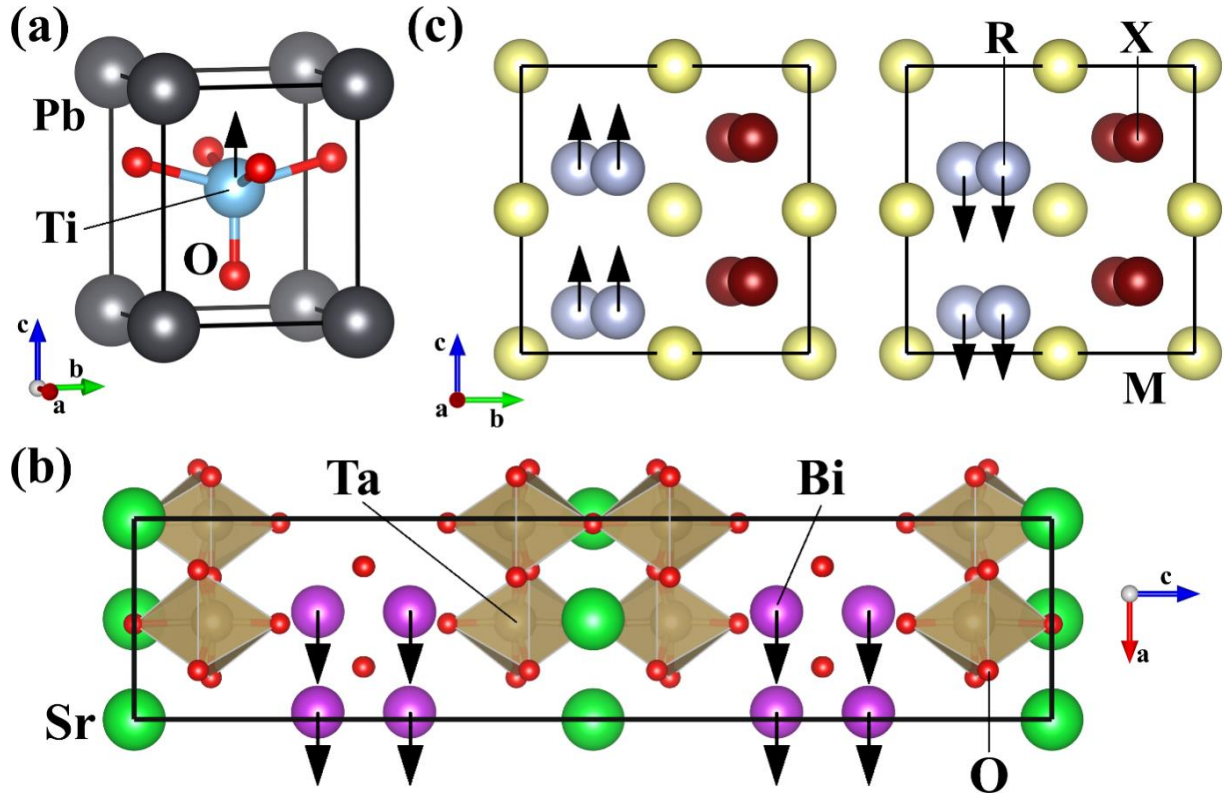

**Figure S3.** Illustration of the major atomic movement direction during polarization reversal of typical ferroelectrics. (a) PbTiO$_3$ in the $P4mm$ structrure; (b) SrBi$_2$Ta$_2$O$_9$ in the $A2_1am$ structure; (c) MRX in the $Pca2_1$ structure, including possible modes in opposite directions.

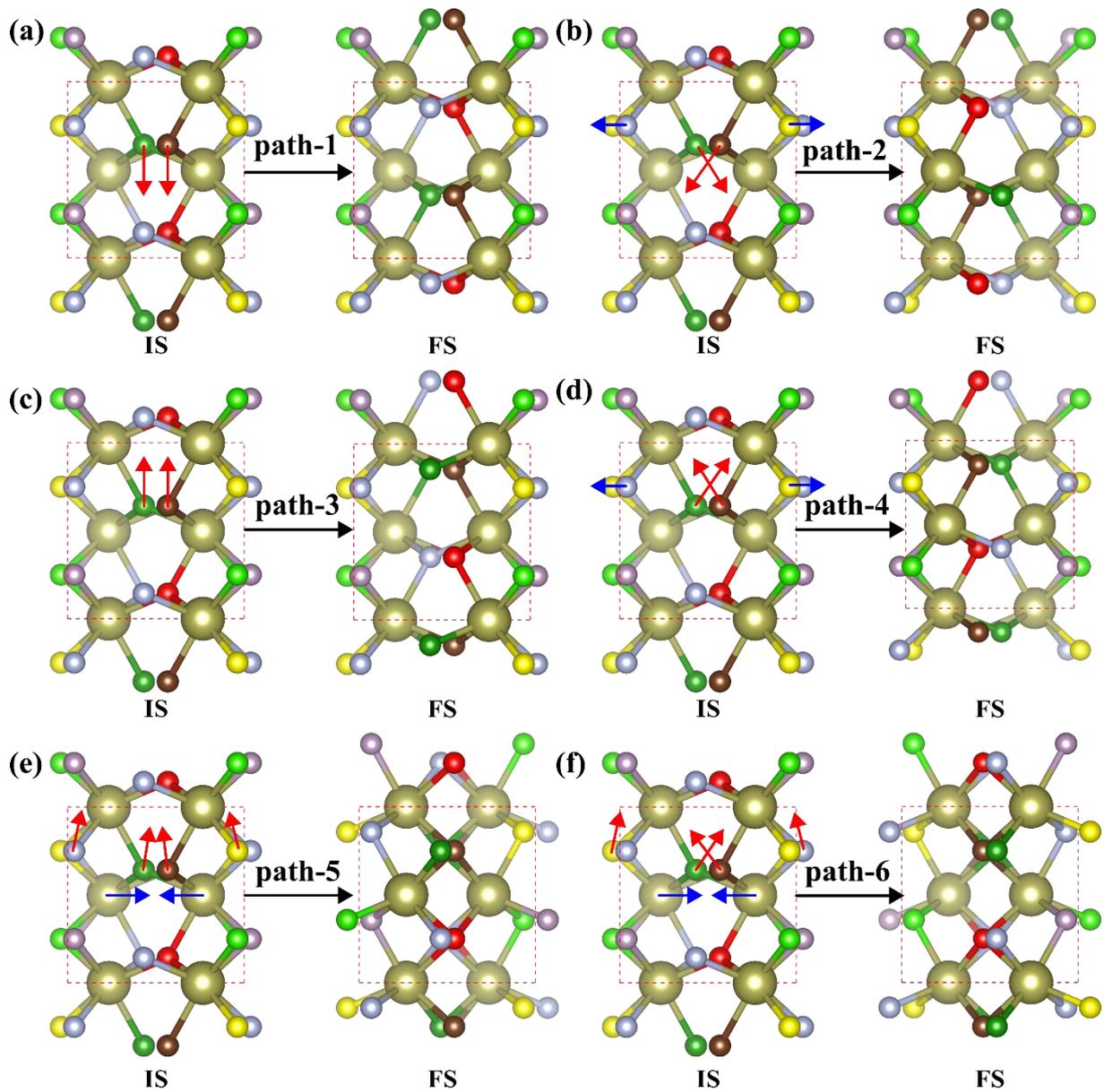

**Figure S4.** Illustration of all six switching paths in *Pca*2$_1$-HfO$_2$. The "IS" and "FS" represent initial state and final state of *Pca*2$_1$-HfO$_2$ in the CI-NEB calculations. The arrows denote the ion migration directions in the switching process. The eight oxygen atoms in the *Pca*2$_1$-HfO$_2$ phase are distinguished by different colors.

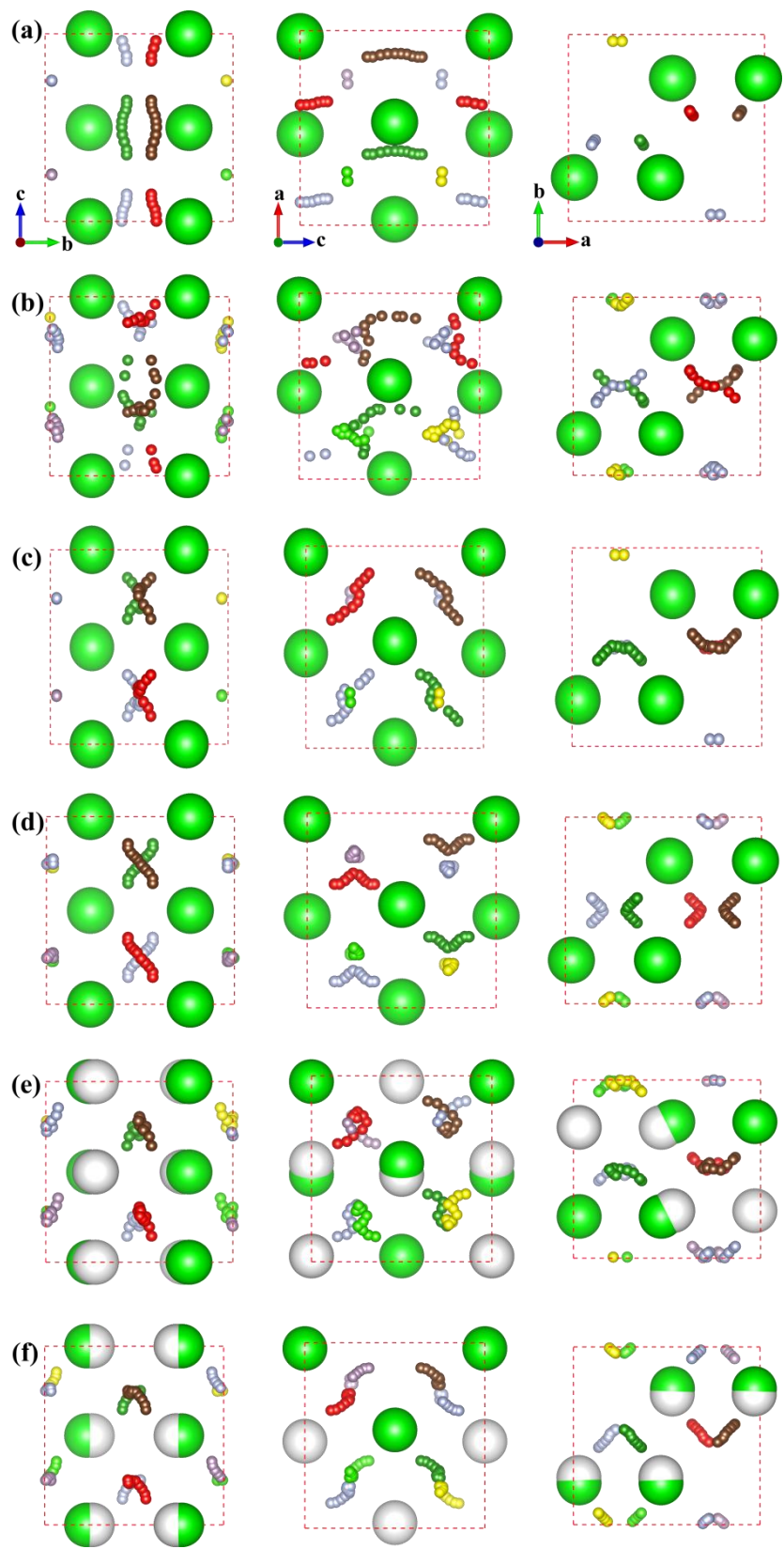

**Figure S5.** Illustration of all six switching paths in $Pca2_1$-$ZrO_2$, viewed from three perspectives.

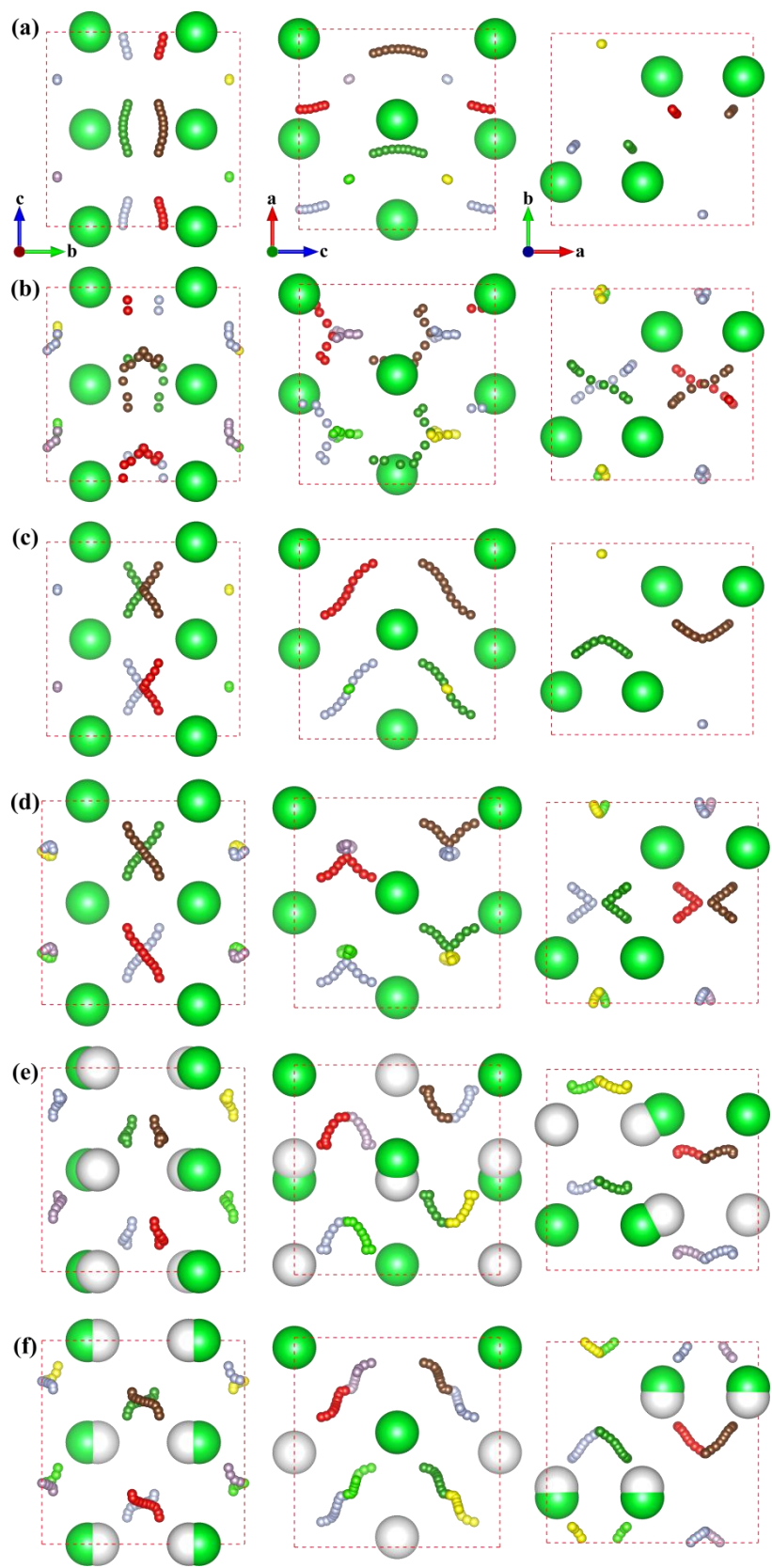

**Figure S6.** Illustration of all six switching paths in $Pca2_1$-$SrI_2$, viewed from three perspectives.

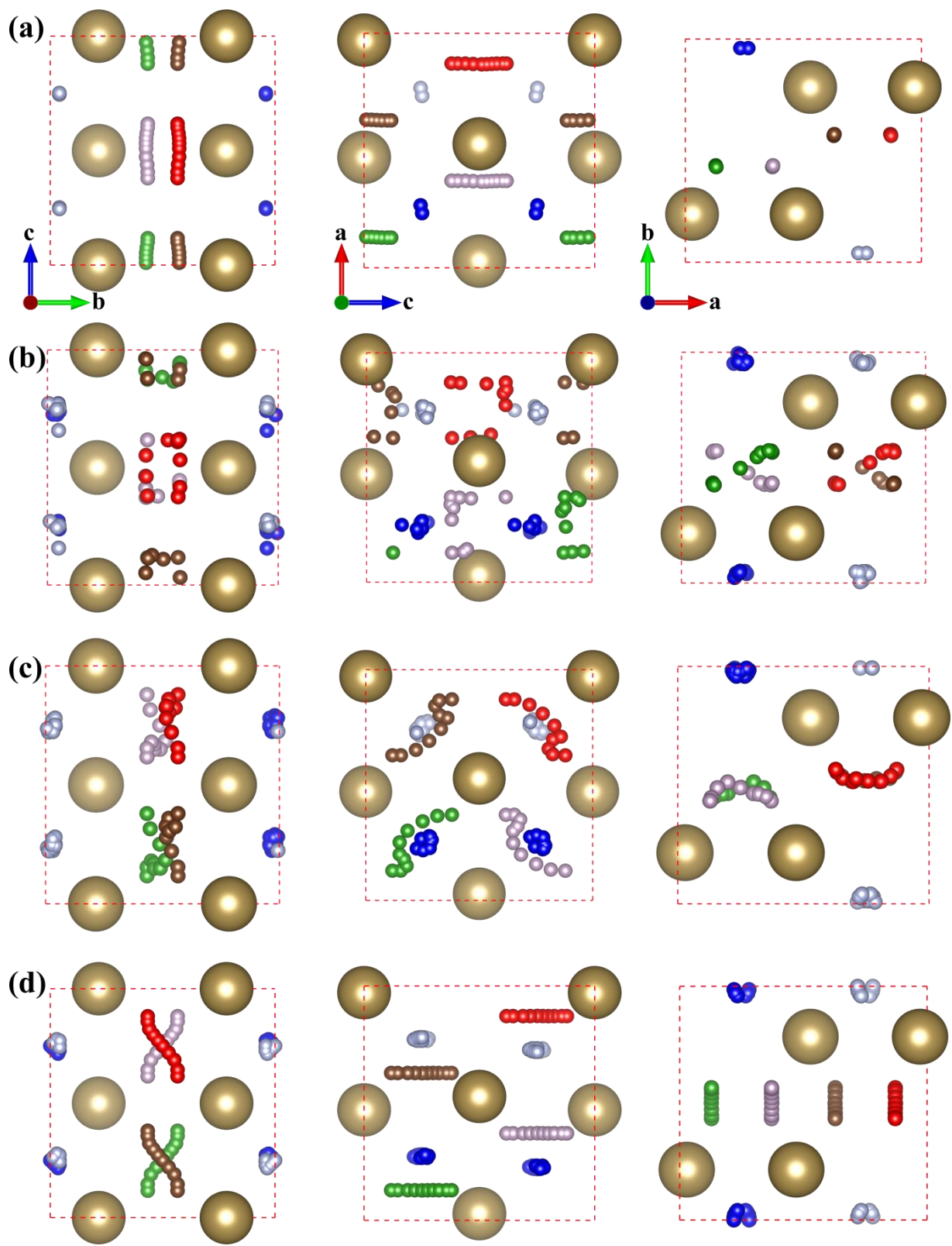

**Figure S7.** Illustration of all four switching paths in *Pca*2$_1$-TaON, viewed from three perspectives.

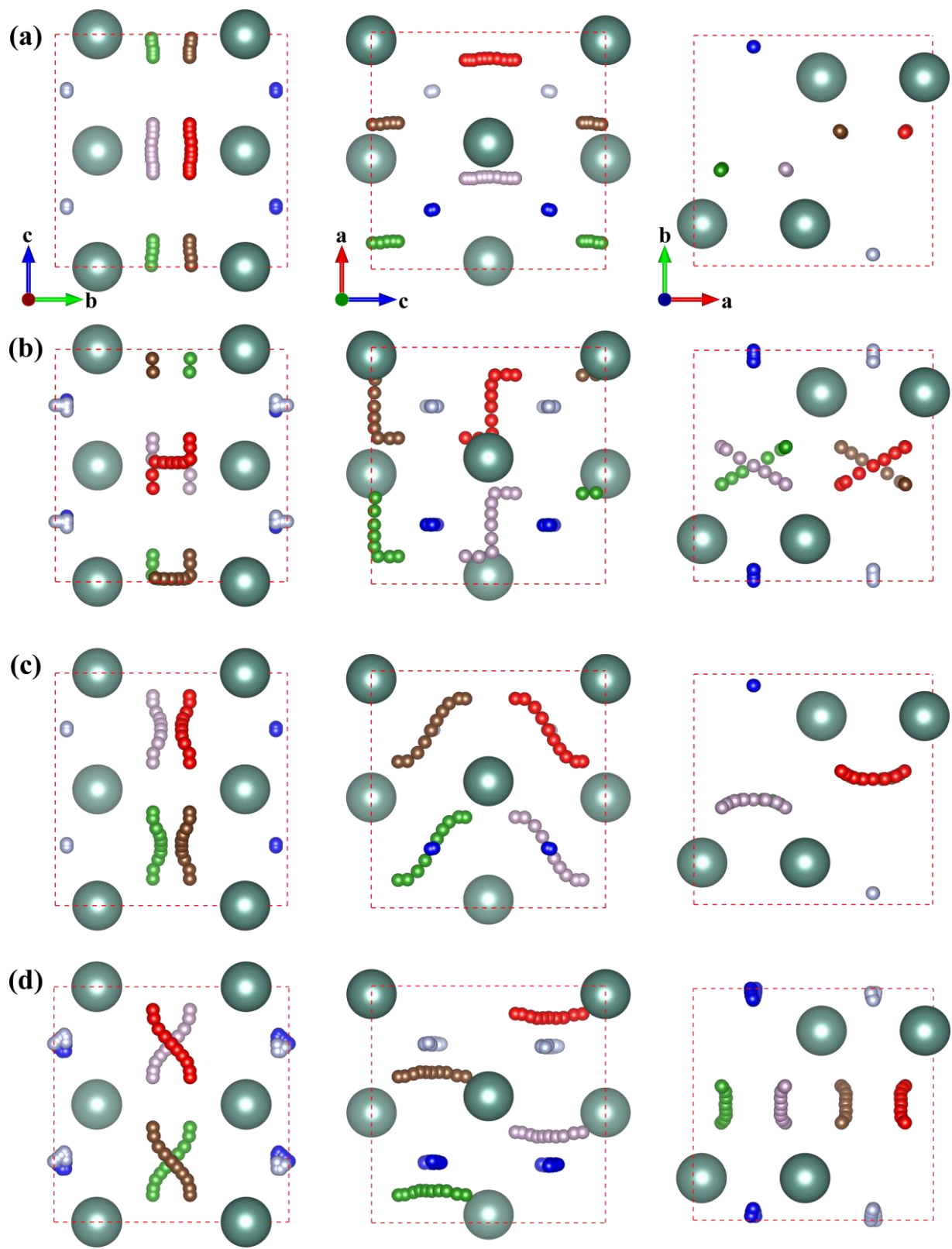

**Figure S8.** Illustration of all four switching paths in *Pca*2$_1$-YSBr, viewed from three perspectives.

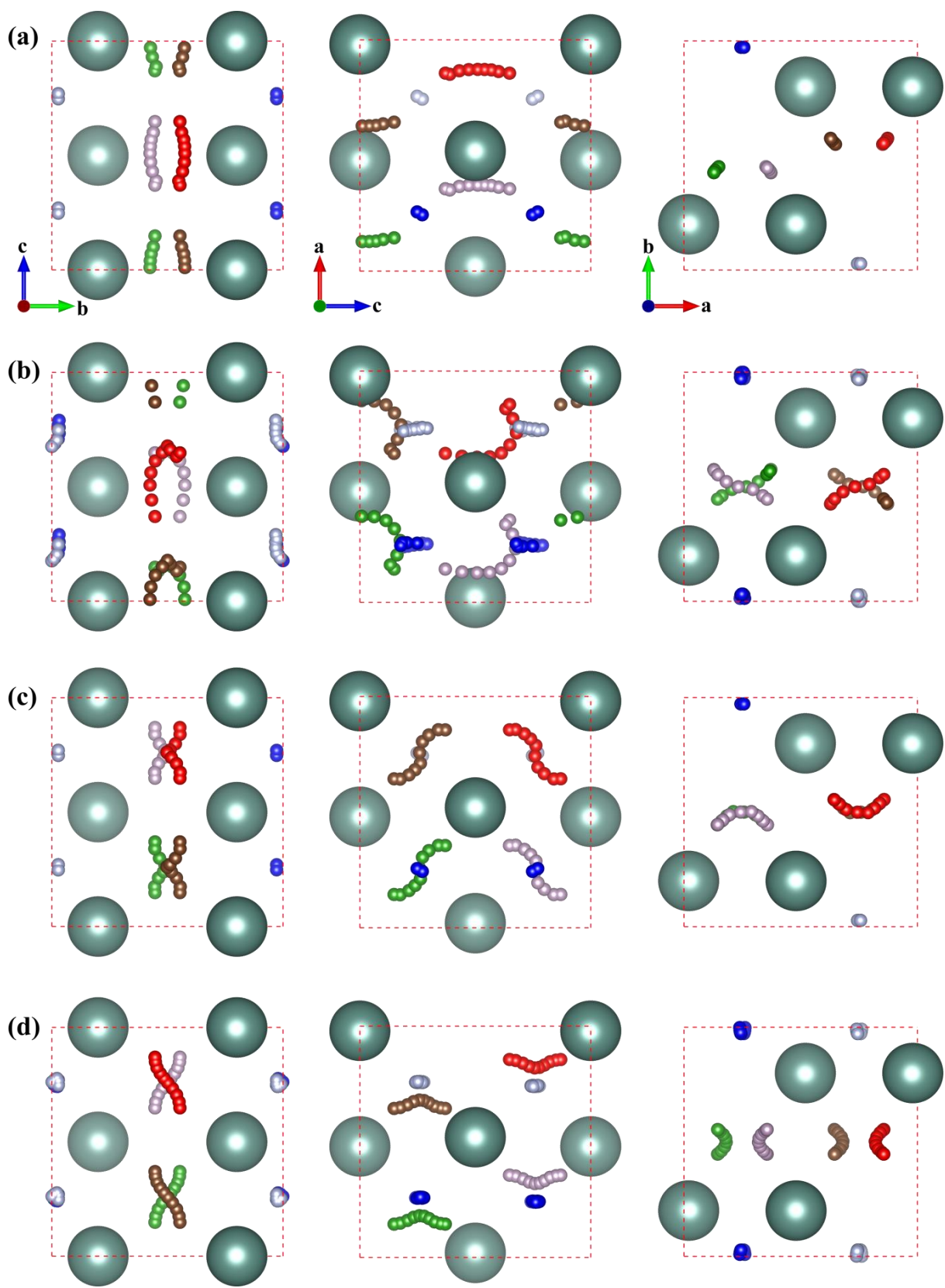

**Figure S9.** Illustration of all four switching paths in *Pca*2₁-YOF, viewed from three perspectives.

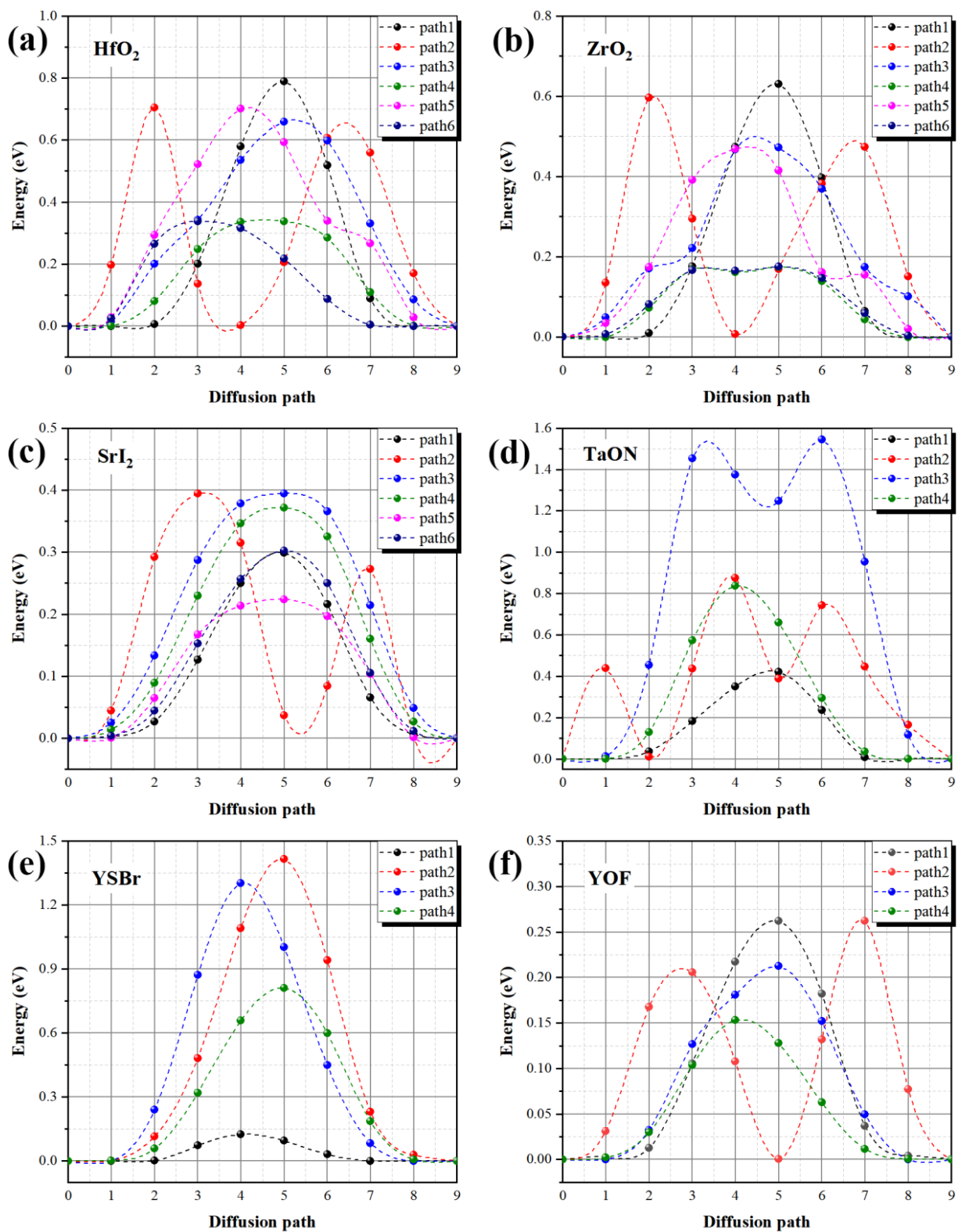

**Figure S10.** CI-NEB results of various paths in *Pca*2$_1$-compounds. (a) HfO$_2$, (b) ZrO$_2$, (c) SrI$_2$, (d) TaON, (e) YSBr, and (f) YOF.

# References for the Supplementary Information